\documentstyle[11pt,twoside,paspconf,epsf]{article}

\begin{document}

\title{Dark Matter and Dark Energy in the Universe}

\author{Michael S. Turner\altaffilmark{1}}
\affil{Departments of Astronomy \& Astrophysics and of Physics,
Enrico Fermi Institute, The University of Chicago}

\altaffiltext{1}{Also, NASA/Fermilab Astrophysics Center,
Fermi National Accelerator Laboratory}

\begin{abstract}

For the first time, we have a plausible, complete accounting
of matter and energy in the Universe.  Expressed a fraction
of the critical density it goes like this:  neutrinos, between 0.3\%
and 15\%; stars, 0.5\%; baryons (total), 5\%; matter (total), 40\%;
smooth, dark energy, 60\%; adding up to the critical density
(summarized in Fig.~\ref{fig:omega}).
This accounting is consistent with the inflationary prediction
of a flat Universe and defines three dark-matter problems:
Where are the dark baryons?  What is the nonbaryonic dark matter?
What is the nature of the dark
energy?  The leading candidate for the (optically) dark baryons
is diffuse hot gas; the leading candidates for the nonbaryonic
dark matter are slowly moving elementary particles left over from
the earliest moments (cold dark matter), such as axions or
neutralinos; the leading candidates for the dark energy involve
fundamental physics and include a
cosmological constant (vacuum energy), a rolling scalar field
(quintessence), and light, frustrated topological defects.

\end{abstract}


\section{Introduction}

The quantity and composition of matter and energy in the
Universe is a fundamental and important issue in cosmology.  The
fraction of the critical energy density contributed by
matter and energy today,
\begin{equation}
\Omega_0 \equiv {\rho_{\rm tot} \over \rho_{\rm crit}}
= \sum_i \Omega_i \,,
\end{equation}
determines the geometry of the Universe:
\begin{equation}
R_{\rm curv}^2 = {H_0^{-2} \over \Omega_0 - 1}\,.
\end{equation}
Here, $\rho_{\rm crit} = 3H_0^2/8\pi G \simeq 1.88h^2\times
10^{-29}\,{\rm g\ cm^{-3}}$, $\Omega_i$ is the fraction of
critical density contributed by component $i$ today (e.g.,
baryons, photons, stars, etc) and $H_0 = 100h\,{\rm km\,s^{-1}\,
Mpc^{-1}}$.

\begin{figure}
\plotone{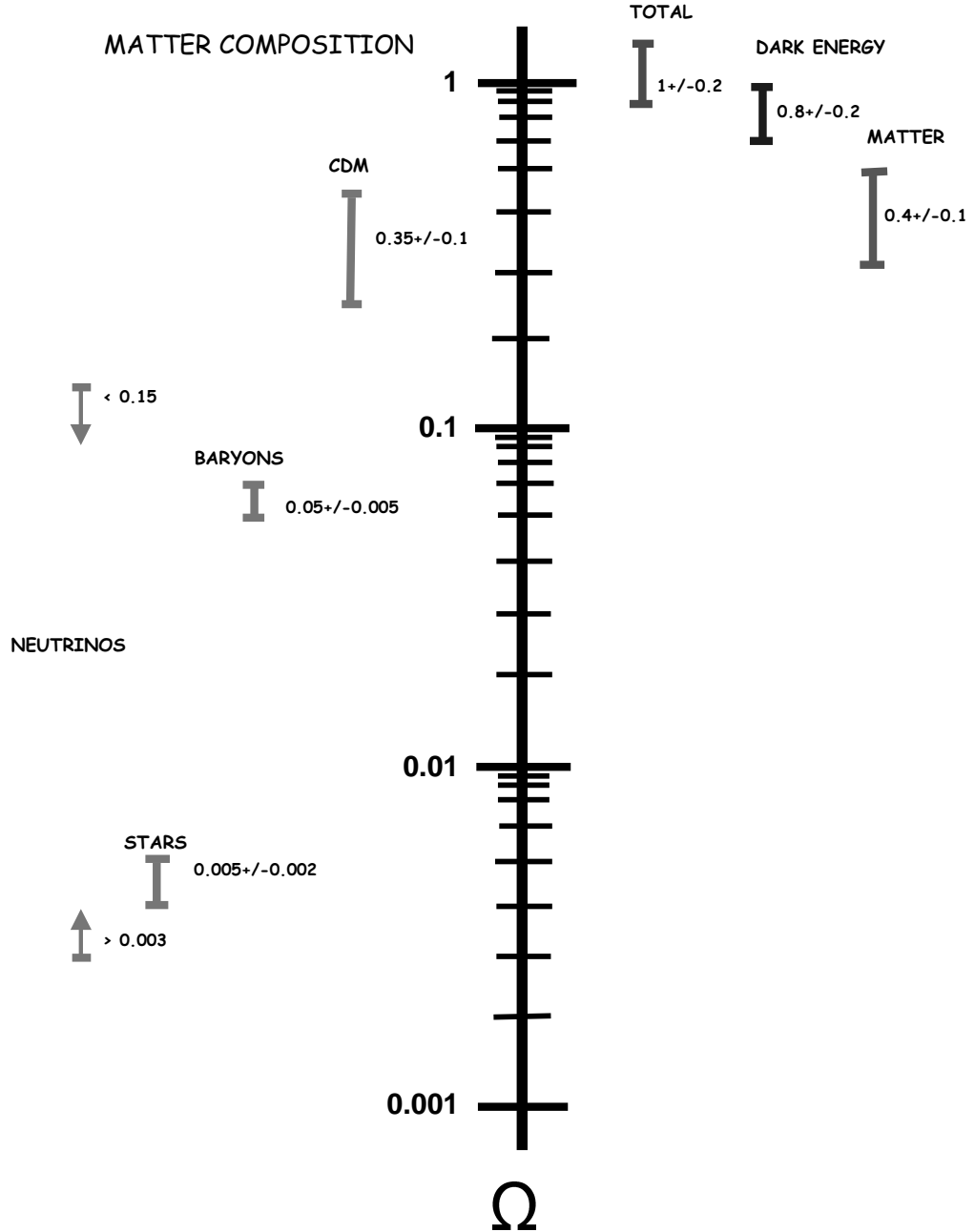}
\caption{Summary of matter/energy in the Universe.
The right side refers to an overall accounting of matter
and energy; the left refers to the composition of the matter
component.  The contribution of relativistic particles,
CBR photons and neutrinos, $\Omega_{\rm rel}h^2 = 4.170\times
10^{-5}$, is not shown.  The upper limit to mass density contributed
by neutrinos is based upon the failure of the hot dark
matter model of structure formation
and the lower limit follows from the
evidence for neutrino oscillations (Fukuda et al, 1998).
Here $H_0$ is taken to be $65\,{\rm km\,s^{-1}\,Mpc^{-1}}$.
}
\label{fig:omega}
\end{figure}

Supplemented by the equation of state for matter and
energy in the Universe, $\Omega_0$ determines the
present rate of deceleration (or acceleration as the
case may be) of the expansion
\begin{equation}
q_0 \equiv {(\ddot R /R)_0 \over H_0^2}
        = {1 \over 2}\Omega_0 + {3\over 2} \sum_i \Omega_i w_i \,,
\end{equation}
where the pressure of component $i$, $p_i \equiv w_i \rho_i$
(e.g., for baryons $w_i = 0$, for radiation $w_i = 1/3$,
and for vacuum energy $w_i = -1$).

The fate of the Universe -- expansion forever or recollapse --
is not directly determined by $H_0$, $\Omega_0$ and $q_0$.
It also depends upon knowing the composition of {\em all}
components of matter and energy for all times in the future.
Recollapse occurs only if there is a future turning point, that is
a future epoch when the expansion rate,
\begin{equation}
H^2 = {8\pi G \over 3} \sum_i \rho_i - {1\over R_{\rm curv}^2}\,,
\end{equation}
becomes zero and $\ddot R < 0$.  In a universe comprised
of matter alone, only a positively curved universe eventually recollapses.
Exotic components can complicate matters:  a positively curved
universe with positive vacuum energy can expand forever, and
a negatively curved universe with negative vacuum energy can recollapse.

The quantity and composition of matter and energy in the Universe
is crucial for understanding the past as well as the future.  It
determines the present age of the Universe, when the Universe ended
its early radiation dominated era, the growth of small inhomogeneities
in the matter and ultimately how large-scale structure formed
in the Universe, as well as the formation and evolution of individual galaxies.

Measuring the quantity and composition of matter and energy in the Universe
is a challenging task.  Not just because the scale of inhomogeneity
is so large, around $10\,$Mpc; but also, because there may be components
that remain exactly or relatively smooth (e.g., vacuum energy or
relativistic particles) which only reveal themselves by their
influence of the evolution of the Universe itself.

Because it is known to be black-body radiation to very high precision
(better than $0.005\%$) and its temperature is known to four
significant figures, $T_0 = 2.7277\pm 0.002\,$K,
the contribution of the cosmic background radiation (CBR)
is very precisely known, $\Omega_\gamma h^2 = 2.480\times 10^{-5}$.
If neutrinos are massless or very light, $m_\nu \ll 10^{-4}\,$eV,
their energy density is equally well known because it is directly
related to that of the photons, $\Omega_\nu = {7\over 8}(4/11)^{4/3}
\Omega_\gamma$ (per species) (actually, there is a 1\% positive
correction to this number; see Dodelson \& Turner, 1992).

The matter component (denoted by $\Omega_M$), i.e., particles
that have negligible pressure, is the easiest to determine because
matter clumps and its gravitational effects are thereby enhanced
(e.g., in rich clusters the matter density typically exceeds
the mean density by a factor of 1000 or more).
With all of this in mind, I will decompose the present matter/energy
density into two components, matter and vacuum energy,
\begin{equation}
\Omega_0 = \Omega_M + \Omega_\Lambda\,.
\end{equation}
I will not again mention the contribution of the CBR
and ultrarelativistic neutrinos and will use vacuum energy
as a stand in for any smooth component (more later).  Vacuum energy
and a cosmological constant are indistinguishable:  a cosmological
constant corresponds to a uniform energy density of magnitude
$\rho_{\rm vac} = \Lambda /8\pi G$.

\section{A Complete Inventory of Matter and Energy}

\subsection{Curvature}

There is a growing consensus that the anisotropy of the CBR
offers the best means of determining $\Omega_0$ and
the curvature of the Universe.  This is because the method is geometric --
standard ruler on the last-scattering surface --
and involves straightforward physics at
a simpler time (see e.g., Kamionkowski et al, 1994).

\begin{figure}
\plotone{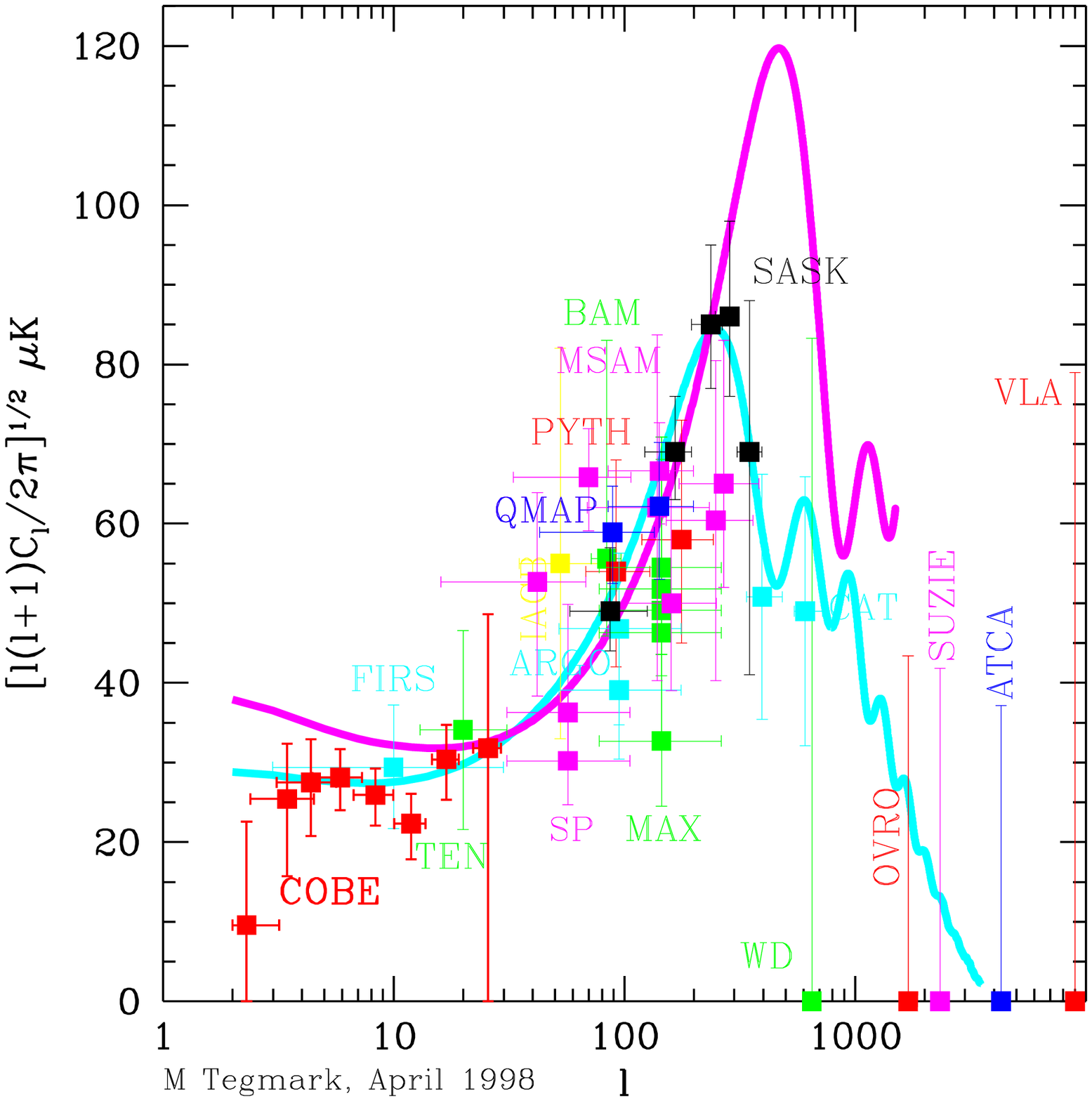}
\caption{Summary of all CBR anisotropy measurements, where
the temperature variations across the sky have been expanded
in spherical harmonics, $\delta T(\theta , \phi ) = \sum_i a_{lm}Y_{lm}$
and $C_l \equiv \langle |a_{lm}|^2\rangle$.  In plain
language, this plot shows the size of the temperature variation
between two points on the sky separated by angle $\theta$
(ordinate) vs. multipole number $l=200^\circ / \theta$
($l=2$ corresponds to $100^\circ$, $l=200$ corresponds to $\theta = 1^\circ$,
and so on).  The curves illustrate the predictions of CDM models
with $\Omega_0 = 1$ (curve with lower peak) and $\Omega_0 =0.3$ (darker
curve).  Note:  the preference of the data for a flat Universe, and
the evidence for the first of a series of ``acoustic peaks.''
The presence of these acoustic peaks is a key signature
of the density perturbations of quantum origin predicted by inflation
(Figure courtesy of M. Tegmark).}
\label{fig:cbr_today}
\end{figure}

At last scattering baryons were still tightly
coupled to photons; as they fell into the dark-matter
potential wells the pressure of photons acted as a restoring
force, and gravity-driven acoustic oscillations resulted.  These
oscillations can be decomposed into their Fourier modes;
Fourier modes with $k\sim l H_0/2$ determine the multipole
amplitudes $a_{lm}$ of CBR anisotropy.  Last scattering
occurs over a short time, and thus the CBR is a snapshot of
the Universe at $t_{\rm ls} \sim 300,000\,$yrs.  Different Fourier
modes are captured at different phases of their oscillation.  (Note,
for the density perturbations predicted by inflation,
all modes the have same initial phase because
all are growing-mode perturbations.)
Modes caught at maximum compression or rarefaction lead
to the largest anisotropy; this results in a series of
acoustic peaks beginning at $l\sim 200$ (see Fig.~\ref{fig:cbr_today}).
The  wavelength of the lowest
frequency acoustic mode that has reached maximum compression,
$\lambda_{\rm max} \sim v_s t_{\rm ls}$, is the standard
ruler on the last-scattering surface.  Both $\lambda_{\rm
max}$ and the distance to the last-scattering surface depend
upon $\Omega_0$, and the position of the first peak $l\simeq
200/\sqrt{\Omega_0}$.  This relationship is insensitive
to the composition of matter and energy in the Universe.

\begin{figure}
\plotone{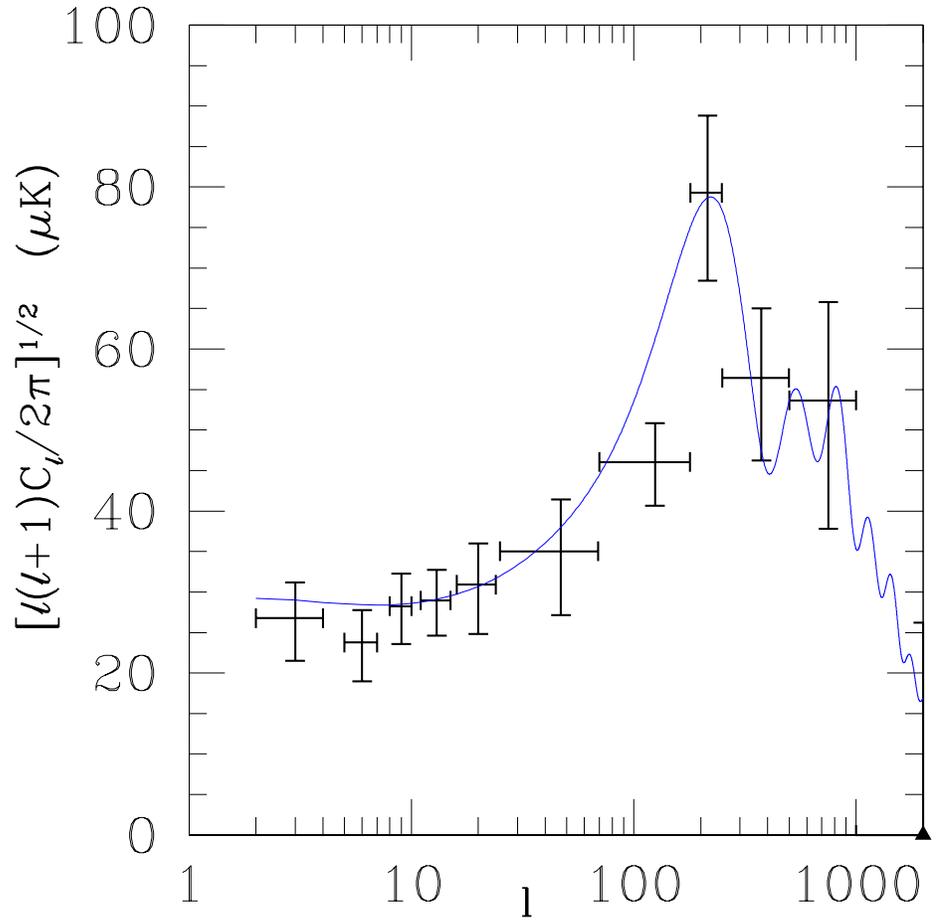}
\caption{The same data as in Fig.~2, but
averaged and binned to reduce error bars and
visual confusion.  The theoretical
curve is for the $\Lambda$CDM model with $H_0=65\,{\rm km\,
s^{-1}\,Mpc^{-1}}$ and $\Omega_M =0.4$ (Figure courtesy of L. Knox).
}
\label{fig:cbr_knox}
\end{figure}

CBR anisotropy measurements, shown in Figs.~\ref{fig:cbr_today} and
\ref{fig:cbr_knox}, now cover
three orders of magnitude in multipole number and come
from more than twenty experiments.  COBE is
the most precise and covers multipoles $l=2-20$;
the other measurements come from balloon-borne, Antarctica-based
and ground-based experiments using both low-frequency
($f<100\,$GHz) HEMT receivers and high-frequency ($f>100\,$GHz)
bolometers.  Taken together, all the measurements are beginning to
define the position of the first acoustic peak, at a value that is
consistent with a flat Universe.  Various analyses of the
extant data have been carried out, indicating $\Omega_0 \sim 1\pm 0.2$
(see e.g., Lineweaver, 1998).
It is certainly too early to draw definite conclusions or put
too much weigh in the error estimate.  However, a strong
case is developing for a flat Universe and more data is on
the way (Maxima, Boomerang, MAT, Python V, DASI, and others).
Ultimately, the issue will be settled by NASA's
MAP (launch late 2000) and ESA's Planck (launch 2007) satellites
which will map the entire CBR sky with 30 times the resolution
of COBE (around $0.1^\circ$).

\newpage

\subsection{Matter}

\begin{figure}
\plotone{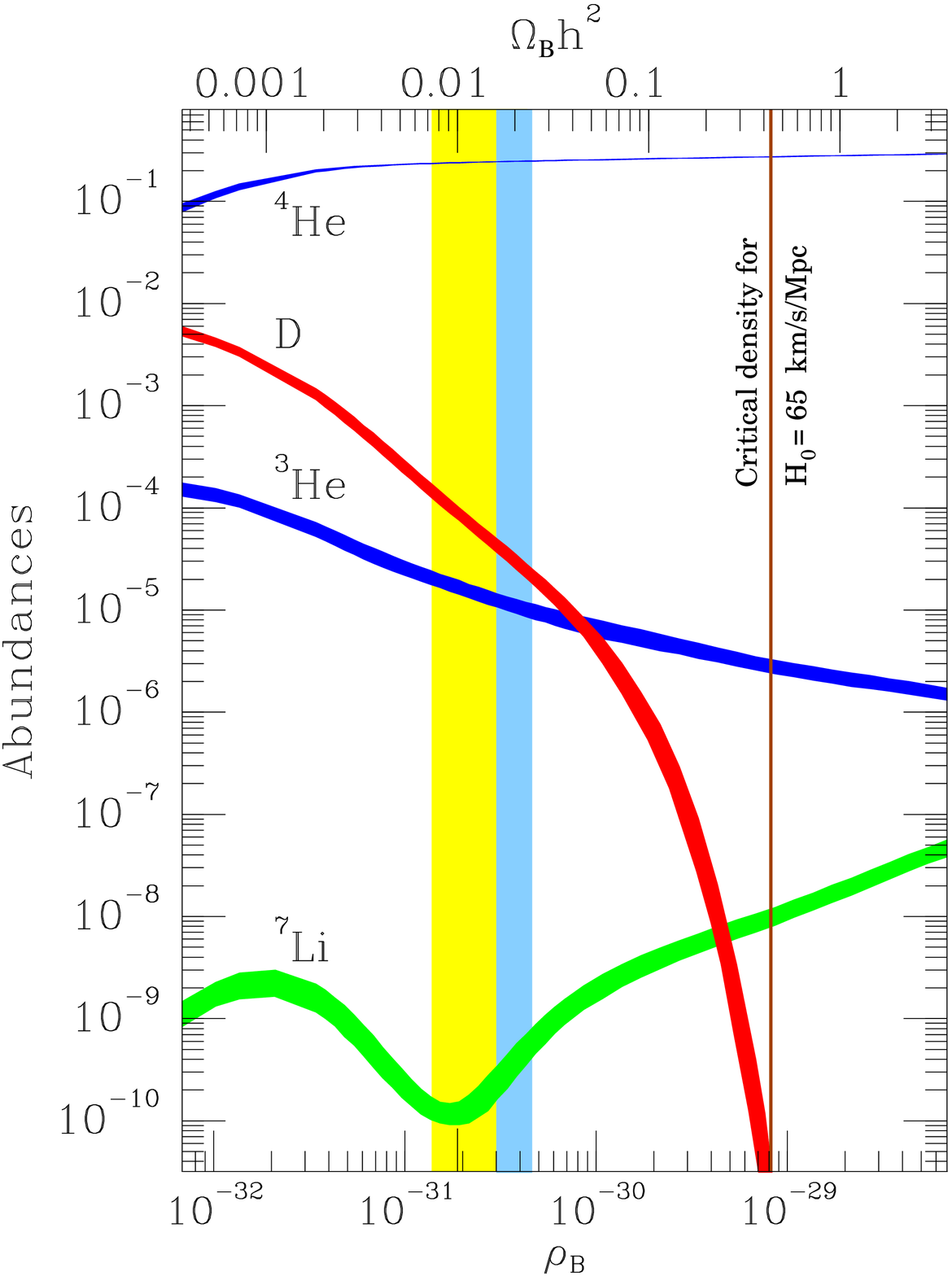}
\caption{Predicted abundances of $^4$He, D, $^3$He, and $^7$Li
(relative to hydrogen) as a function of the density of ordinary
matter (baryons); the width of the curves indicate the ``$2\sigma$''
theoretical uncertainty.
The full vertical band denotes the concordance interval based upon all
four light elements that dates back to 1995 (Copi et al, 1995).
The darker portion highlights
the determination of the baryon density based upon
the recent measurement of the primordial abundance of
deuterium (Burles \& Tytler, 1998a,b), which implies
$\Omega_Bh^2 = 0.02 \pm 0.002$.
}
\label{fig:bbn}
\end{figure}
\subsubsection{Baryons}

For more than twenty years big-bang nucleosynthesis (BBN) has provided
a key test of the hot big-bang cosmology as well as
the most precise determination of the baryon density.
Careful comparison of the primeval abundances of D, $^3$He, $^4$He
and $^7$Li with their big-bang predictions defined a
concordance interval, $\Omega_Bh^2 = 0.007 - 0.024$ (see e.g.,
Copi et al, 1995).

Of the four light elements produced in the big bang,
deuterium is the most powerful ``baryometer'' -- its primeval abundance
depends strongly on the baryon density ($\propto 1/\rho_B^{1.7}$) --
and its the evolution of its abundance since the big bang is simple --
astrophysical processes only destroy deuterium.  Until recently
deuterium could not be exploited as a baryometer because its
abundance was only known locally, where roughly half of the material
has been through stars with a similar amount of
the primordial deuterium destroyed.  In 1998, the situation changed
dramatically.

Over the past four years there have been many claims for upper limits,
lower limits, and determinations
of the primeval deuterium abundance, ranging from (D/H)\,$=10^{-5}$
to (D/H)\,$=3\times 10^{-4}$.  Within the past year
Burles and Tytler have clarified the situation and established
a strong case for (D/H)$_P = (3.4\pm 0.3)\times 10^{-5}$.
Their case is based upon the deuterium abundance measured
in four high-redshift hydrogen clouds seen in absorption against distant QSOs,
and the remeasurement and reanalysis of other putative deuterium systems.
In this important enterprise, the Keck I and its HiRes Echelle
Spectrograph have played the crucial role.
The primordial deuterium measurement turns the previous factor of three
concordance range for the baryon density into
a 10\% determination of the baryon density, $\rho_B =
(3.8\pm 0.4)\times 10^{-31}\,{\rm g\,cm^{-3}}$ or $\Omega_Bh^2 =
0.02 \pm 0.002$ (see Fig.~\ref{fig:bbn}).

It is nice to see that this very precise determination of
the baryon density, based upon the early Universe physics
of BBN, is consistent with two other measures of the baryon
density, based upon entirely different physics.
By comparing measurements of the opacity of the Lyman-$\alpha$
forest toward high-redshift quasars with high-resolution, hydrodynamical
simulations of structure formation, several groups (Meiksin \& Madau, 1993;
Rauch et al, 1997; Weinberg et al, 1997)
have inferred a lower limit to the baryon density,
$\Omega_Bh^2 > 0.015$ (it is a lower limit because it depends
upon the baryon density squared divided by the
intensity of the ionizing radiation field).
The second test involves the height of the first acoustic peak:
it rises with the baryon density (the higher the baryon density,
the stronger the gravitational force driving the acoustic
oscillations).  Current CBR measurements are consistent with
the Burles -- Tytler baryon density; the MAP and
Planck satellites should ultimately provide a 5\% or better
determination of the baryon density, based upon the
physics of gravity-driven
acoustic oscillations when the Universe was 300,000\,yrs old.
This will be an important cross check of the BBN determination.

\subsubsection{It's dark!}

Based upon the mass-to-light ratios of the bright, inner regions of
galaxies and the luminosity density of the Universe, the fraction
of critical density in stars has been determined,
\begin{equation}
\Omega_* = (M/L)_* {\cal L} /\rho_{\rm crit} =
(M/L)_* / 1200h \simeq (0.003\pm 0.001)h^{-1}\,.
\end{equation}
Since the pioneering work of
Fritz Zwicky and Vera Rubin, it has been known that this is far too little
material to hold galaxies and clusters together, and thus, that
most of the matter in the Universe is dark.  Determining the
total amount of dark matter has been the challenge.  At present,
I believe that clusters seem to provide the most reliable estimate
of the matter density.

\begin{figure}
\plotone{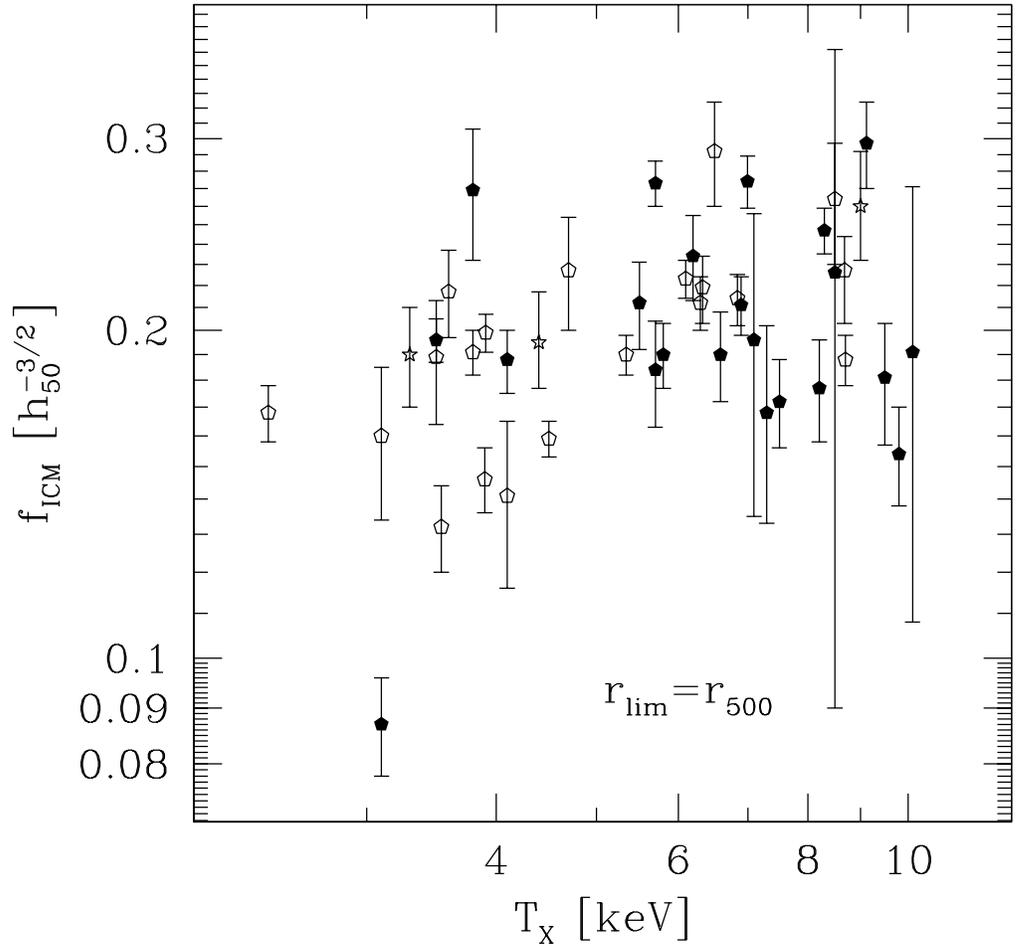}
\caption{Cluster gas fraction as a function of
cluster gas temperature for a sample of 45 galaxy clusters
(Mohr et al, 1998).  While there is some indication
that the gas fraction
decreases with temperature for $T< 5\,$keV, perhaps because
these lower-mass clusters lose some of their hot gas, the
data indicate that the gas fraction reaches a plateau
at high temperatures, $f_{\rm gas} =0.075 \pm 0.002$.
}
\label{fig:gas}
\end{figure}

\subsubsection{Total density of matter}

Rich clusters formed from density perturbations of size
around 10\,Mpc; in so doing they gather a sample
of matter from a very large region, large enough to provide
a ``fair sample'' of matter in the Universe.
Using clusters as such,
the precise BBN baryon density can be used
to infer the total matter density (White et al, 1993).
(Note, the baryons and dark matter need not be
well mixed, provided that the baryon and total mass are determined
over a large enough region of the cluster.)

Most of the baryons in clusters reside in the
hot, x-ray emitting intracluster gas and not in the galaxies
themselves,
and so the problem reduces to determining the gas-to-total mass ratio.
The gas mass can be determined by two methods:
1) measuring the x-ray flux from the intracluster
gas and 2) mapping the Sunyaev - Zel'dovich
CBR decrement caused by CBR photons scattering off hot electrons in the
intracluster gas.  The total cluster mass can be determined
three independent ways:  1)  using the motion of the galaxies
and the virial theorem; 2) assuming that the gas is in hydrostatic
equilibrium and using the virial theorem for the gas; and
3) mapping the cluster mass by gravitational lensing.  Within their
uncertainties and where comparisons can be made, the three methods
for determining the total mass agree; likewise, the two methods
for determining the gas mass are consistent.

Mohr et al (1998) have compiled the gas to total mass ratios determined from
x-ray measurements for a sample of 45 clusters; they find
$f_{\rm gas} = (0.07\pm 0.002)h^{-3/2}$.  Carlstrom (1999), using his
S-Z gas measurements and x-ray measurements for the total
mass for 27 clusters, finds $f_{\rm gas} =(0.06\pm 0.006)h^{-1}$.
(The agreement of these two numbers means that clumping of
the gas, which could lead to an overestimate of the gas fraction
based upon the x-ray flux, is not a problem.)
Using the fair sample assumption, the mean
matter density in the Universe can be inferred:
\begin{eqnarray}
\Omega_M = \Omega_B/f_g & = & (0.3\pm 0.05)h^{-1/2}\  ({\rm X ray})\nonumber\\
                   & = &  (0.25\pm 0.04)h^{-1}\ ({\rm S-Z}) \nonumber \\
                   & = & 0.4\pm 0.1\ ({\rm my\ summary})\,.
\end{eqnarray}
At present, I believe this to be the most reliable and precise
determination of the matter density.  It involves few assumptions,
and most of them have now been tested (clumping, hydrostatic equilibrium,
variation of gas fraction with cluster mass).

\subsubsection{Supporting evidence for $\Omega_M=0.4\pm 0.1$}
This result is consistent with
a variety of other methods, that involve very different physics.
For example, based upon the evolution
of the abundance of rich clusters with redshift, Henry (1998)
finds $\Omega_M
= 0.45 \pm 0.1$ (also see, Bahcall \& Fan, 1998).
Dekel and Rees (1994) place a low limit $\Omega_M > 0.3$ (95\% cl)
derived from the outflow of material from voids (a void
effectively acts as a negative mass proportional to the
mean matter density).

The analysis of the peculiar velocities of
galaxies provides an important probe of the mass density averaged
over very large scales (of order several 100\,Mpc).  By comparing
measured peculiar velocities with those predicted from the
distribution of matter determined by redshift surveys such as
the IRAS survey of infrared galaxies,
one can infer the quantity $\beta = \Omega_M^{0.6}/b_I$
where $b_I$ is the linear bias factor that relates the inhomogeneity
in the distribution of IRAS galaxies to that in the distribution of matter
(in general, the bias factor is expected to be in the
range 0.7 to 1.5; IRAS galaxies are expected to be less
biased.).  Recent work by
Willick \& Strauss (1998) finds $\beta = 0.5\pm 0.05$, while Sigad et al
(1998) find $\beta = 0.9\pm 0.1$.  The apparent inconsistency of
these two results and the ambiguity introduced by bias
preclude a definitive determination
of $\Omega_M$ by this method.  However, Dekel (1994) quotes a 95\%
confidence lower bound, $\Omega_M >0.3$, and the work of Willick
\& Strauss seems to strongly indicate that $\Omega_M$ is much less than 1.

Finally, there is strong, but circumstantial, evidence from structure formation
that $\Omega_M$ is around 0.4 and significantly greater than $\Omega_B$.
Since the demise of Peebles' isocurvature baryon model (Peebles, 1987)
some five years ago due to its prediction of excessive CBR
anisotropy on small angular scales,
there has been no model for structure formation without
nonbaryonic dark matter.  The basic reason is simple:  in a baryons only
model, density perturbations only grow from decoupling, $z\sim
1000$, until the Universe becomes curvature dominated, $z\sim \Omega_B^{-1}
\sim 20$; this is simply not enough growth to produce all the structure
seen today with the size of density perturbations inferred from
CBR anisotropy.  With nonbaryonic dark matter, dark matter perturbations
begin growing much earlier and grow until the present epoch, or nearly so.

In addition, the transition from radiation domination at early
times to matter domination
determines the shape of the present power spectrum of density
perturbations, with the redshift of matter -- radiation equality
depending upon $\Omega_Mh^2$.
Measurements of the shape of the present power spectrum based upon redshift
surveys indicate that the shape parameter, $\Gamma = \Omega_M h
\sim 0.25\pm 0.05$ (see e.g., Peacock \& Dodds, 1994).  For
$h\sim 2/3$, this implies $\Omega_M \sim 0.4$.
(If there are relativistic particles beyond the CBR photons
and relic neutrinos, the formula for the shape parameter changes and
$\Omega_M\sim 1$ can be accommodated; see Dodelson et al, 1996).

\subsection{Mass-to-light ratios:  the glass is half full!}

The most mature approach to estimating the matter
density involves the use of mass-to-light ratios, the measured
luminosity density, and the simple equality
\begin{equation}
\langle \rho_M \rangle = \langle M/L \rangle \, {\cal L}\,,
\end{equation}
where ${\cal L}= 2.4h\times 10^8\,L_{B\odot}\,
{\rm Mpc^{-3}}$ is the luminosity density of the Universe.
Once the average mass-to-light ratio for
the Universe is determined, $\Omega_M$ follows
by dividing it by the critical mass-to-light ratio,
$(M/L)_{\rm crit} = 1200h$ (in solar units).  Though it is
tantalizingly simple -- and it is far too easy to take any measured
mass-to-light ratio and divide it by $1200h$ -- this method
does not provide an easy and reliable method of determining $\Omega_M$.

The CNOC group (Carlberg et al, 1996, 1997) have done a
very careful job of determining a mean cluster mass-to-light
ratio, $(M/L)_{\rm cluster} = 240\pm 50$, which translates
to an estimate of the mean matter density,
$\Omega_{\rm cluster} = 0.20 \pm 0.04$.
Because clusters contain thousands of galaxies and cluster
galaxies do not seem {\em radically} different from field galaxies,
one is tempted to take this estimate of the mean matter density
very seriously.  However, it is significantly smaller than the
value I advocated earlier, $\Omega_M = 0.4\pm 0.1$.  Which
estimate is right?

I believe the higher number, based upon the cluster baryon
fraction, is correct and that we should be surprised that
the CNOC number is so close, closer than we had any right to expect!
After all, only a small fraction of galaxies
are found in clusters and the luminosity density ${\cal L}$ itself
evolves strongly with redshift and corrections for this effect
are large and uncertain.  (We are on the tail end of star formation
in the Universe:  80\% of star formation took place
at a redshift greater than unity.)

Even if mass-to-light ratios were measured
in the red (they typically are not),
where the starlight is dominated by
low-mass stars and reflects
the integrated history of star formation
rather than the present rate as blue light does,
one would still require the fraction of baryons converted into stars in
clusters to be identical to that in the field to have agreement
between the CNOC estimate and that based upon the cluster baryon
fraction.  Apparently, the fraction of baryons converted into
stars in the field and in clusters is similar, but not identical.

To put this in perspective and to emphasize the shortcomings
of the mass-to-light technique, had one used the cluster mass-to-x-ray ratio
and the x-ray luminosity density, one would have inferred $\Omega_M
\sim 0.05$.  A factor of two discrepancy based upon this
method is not so bad.  Enough said.


\subsection{Missing energy found!}

The results $\Omega_0 = 1\pm 0.2$ and $\Omega_M = 0.4\pm 0.1$
are in apparent conflict.  However, prompted by a strong
belief in a flat Universe, theorists have explored the logical possibility
a dark, exotic form of energy that is smoothly distributed and
contributes 60\% of the critical density to explain this
discrepancy (Turner et al, 1984; Peebles, 1984).
To avoid interfering with structure formation, this energy density
must be less important in the past than it is today (development
of the structure observed today from density perturbations of the
size inferred from measurements of the anisotropy of the CBR
requires that the Universe be matter dominated from the epoch
of matter -- radiation equality until very recently).  If the
effective equation of state for this component is parameterized as
$w_x = p_X/\rho_X$, its energy density evolves as $\rho_X \propto
R^{-n}$ where $n=3(1+w_x)$.  To be less important in the past than
matter, $n$ must be less than $3$ or $w_X< 0$; the more negative
$w_X$ is, the faster this component gets out of the way.  Another
added benefit of negative pressure is an older Universe for a
given Hubble constant:  $H_0t_0$ increases with decreasing $w_X$.
The simplest example of an exotic smooth component is vacuum energy,
which is characterized $w_X=-1$.

The ``smoking-gun'' signature of such a smooth component is accelerated
expansion (due to negative pressure); for a cosmological constant,
$q_0 = {1\over 2} \Omega_M - \Omega_\Lambda \sim -0.4$.  This year,
evidence for this smoking gun was presented in the form
of the magnitude -- redshift (Hubble)
diagram for fifty-some SNeIa out to redshifts of nearly 1 (Riess et al, 1998;
Perlmutter et al, 1998).  These two groups, working independently
both found evidence for accelerated expansion.
Perlmutter et al (1998) summarize their results as
\begin{equation}
\Omega_\Lambda = {4\over 3}\Omega_M +{1\over 3} \pm {1\over 6}\,,
\end{equation}
which for $\Omega_M\sim 0.4 \pm 0.1$ implies $\Omega_\Lambda = 0.85 \pm 0.2$,
or just what is needed to explain the missing energy!
(see Fig.~\ref{fig:omegalambda}).

(To explain their startling result in simple terms:  If the distances and
velocities to distant galaxies were all measured at the present, they would
obey a perfect Hubble law, $v=H_0d$, because the expansion of the
Universe is just a (conformal) scaling up of all distances.  However, we see
distant galaxies at an earlier time and so if the expansion is
slowing, their velocities should fall above the Hubble-law prediction;
these two groups found the opposite, implying that the expansion
rate is speeding up.)

The statistical errors reported by the two groups are smaller
than possible systematic errors.  Thus, the believability of
the SNeIa result turn on the reliability
of SNeIa as one-parameter standard candles.  SNeIa are thought
to be associated with the nuclear detonation of Chandrasekhar
mass white dwarfs.  The one parameter is the rate of decline
of the light curve:  The brighter ones decline more slowly (the
so-called Phillips relation).
The lack of a good theoretical understanding of this
(e.g., what is the physical parameter?) is offset
by strong empirical evidence for the relationship
between peak brightness and rate of decline, based upon a sample
of thirty-some nearby SNeIa.  It is reassuring that in all respects
studied, the distant sample of SNeIa appear to be similar to
the nearby sample.  For example, distribution of decline rates
and dispersion about the Phillips relationship.  Further, the local
sample spans a wide range of metallicity, both suggesting that
metallicity is not an important second parameter and most likely
spanning the range of metallicities of the distant sample.  At this
point, it seems fair to say that if there is a problem with SNeIa
as standard candles, it must be subtle.

Riess et al (1998) and Perlmutter et al (1998) have presented a strong case
for accelerated expansion:  their data are impressive and both
groups have been very careful and self-critical.
Cosmologists are even more inclined to believe the SNeIa results
because of the preexisting evidence for a ``missing-energy component,''
which predicted accelerated expansion.

\subsection{Cosmic concordance}

\begin{figure}
\plotone{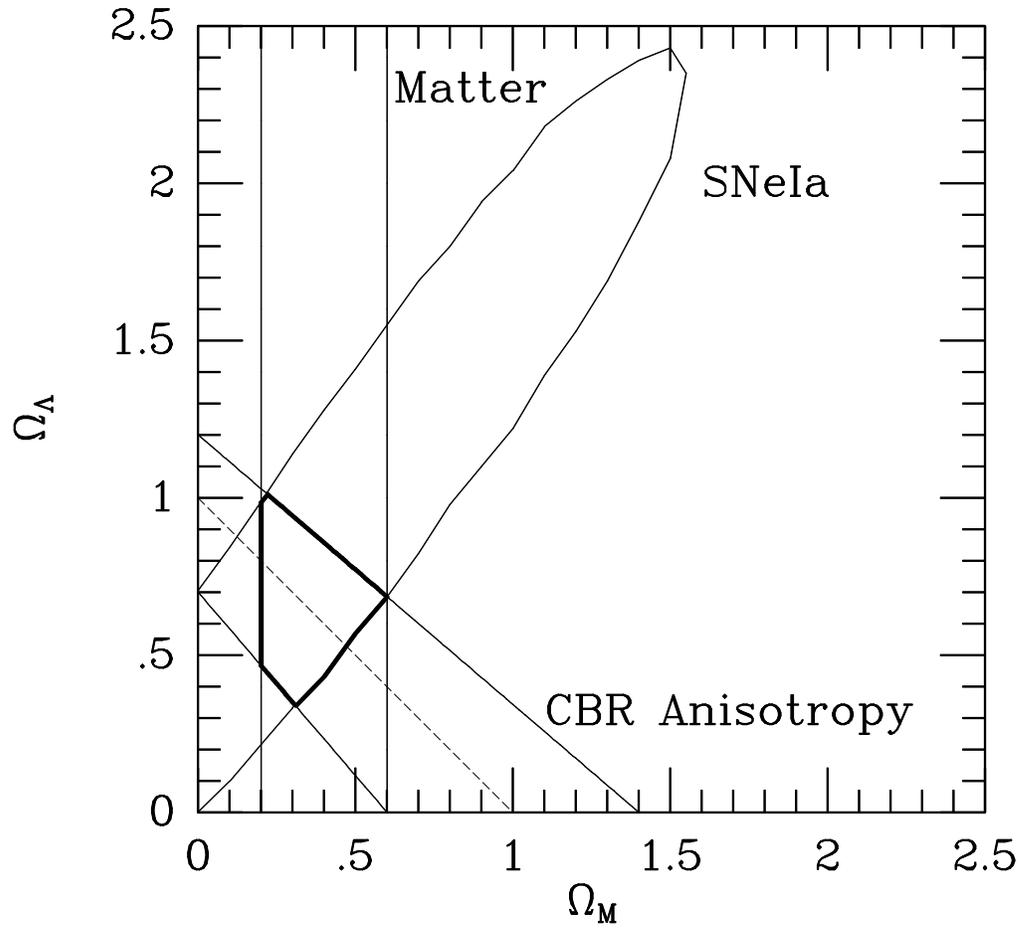}
\caption{Constraints to $\Omega_M$ and $\Omega_\Lambda$
from CBR anisotropy, SNeIa, and measurements of clustered matter.
Lines of constant $\Omega_0$ are diagonal, with a flat
Universe shown by the broken line.
The concordance region is shown in bold:  $\Omega_M\sim 1/3$,
$\Omega_\Lambda \sim 2/3$, and $\Omega_0 \sim 1$.
(Particle physicists who rotate the figure by $90^\circ$
will recognize the similarity to the convergence of the
gauge coupling constants.)
}
\label{fig:omegalambda}
\end{figure}

The reason for my enthusiasm about the SNeIa results is that
for the first time we have a complete, self-consistent
accounting of mass and energy in the Universe, as well as
a self-consistent picture for structure formation.
The consistency of the matter/energy accounting
is illustrated in Fig.~\ref{fig:omegalambda}.
Let me explain this very exciting figure in words.
The SNeIa results are sensitive to the
acceleration (or deceleration) of the expansion, and the results
constrain the combination ${4\over 3}\Omega_M -\Omega_\Lambda$.  (Note,
$q_0 = {1\over 2}\Omega_M - \Omega_\Lambda$; ${4\over 3}\Omega_M -
\Omega_\Lambda$ corresponds to the deceleration parameter
at redshift $z\sim 0.4$, the median redshift of these
samples).  The (approximately) orthogonal combination,
$\Omega_0 = \Omega_M + \Omega_\Lambda$
is constrained by CBR anisotropy.  Together, they define a concordance
region around $\Omega_0\sim 1$, $\Omega_M \sim 1/3$,
and $\Omega_\Lambda \sim 2/3$.  The constraint
to the matter density alone, $\Omega_M = 0.4\pm 0.1$,
provides a cross check, and it is consistent with the previous numbers.
Cosmic concordance!

\begin{figure}
\plotone{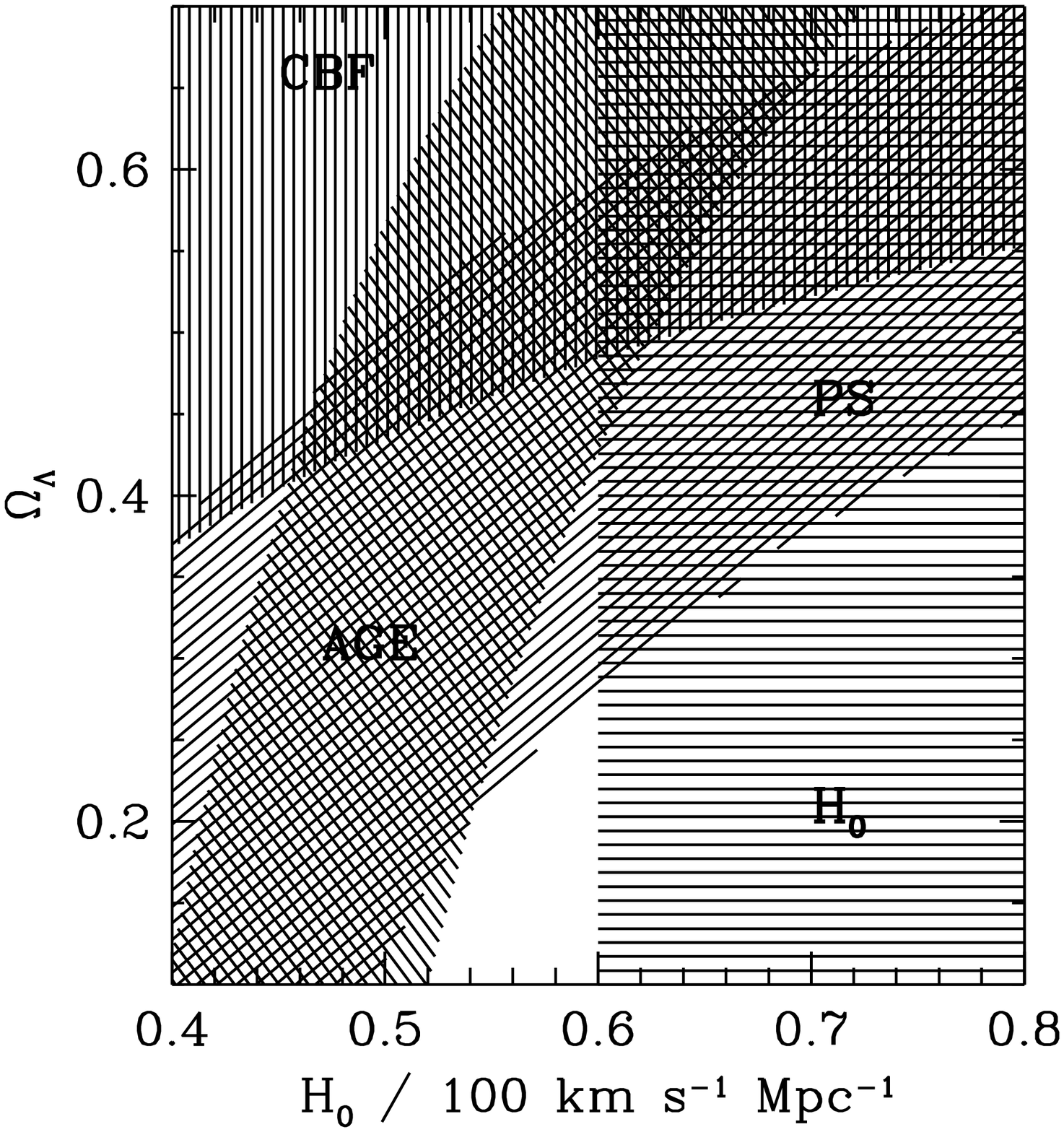}
\caption{Constraints used to determine the best-fit CDM model:
PS = large-scale structure + CBR anisotropy; AGE = age of the
Universe; CBF = cluster-baryon fraction; and $H_0$= Hubble
constant measurements.  The best-fit model, indicated by
the darkest region, has $H_0\simeq 60-65\,{\rm km\,s^{-1}
\,Mpc^{-1}}$ and $\Omega_\Lambda
\simeq 0.55 - 0.65$.  Evidence for its smoking gun signature -- an
accelerating expansion -- was presented in 1998 by Perlmutter
et al and Riess et al.}
\label{fig:best_fit}
\end{figure}

But there is more.  The $\Lambda$CDM model, that is the cold dark
matter model with $\Omega_B \sim 0.05$, $\Omega_{\rm CDM}\sim
0.35$ and $\Omega_\Lambda \sim 0.6$, is a very good fit to all
cosmological constraints:  large-scale structure, CBR anisotropy,
age of the Universe, Hubble constant and the constraints
to the matter density and cosmological constant; see Fig.~\ref{fig:best_fit}
(Krauss \& Turner, 1995; Ostriker \& Steinhardt, 1995;
Turner, 1997).  Further, as
can be seen in Figs.~\ref{fig:cbr_today} and \ref{fig:cbr_knox},
CBR anisotropy measurements are beginning to show evidence for
the acoustic peaks characteristic of the Gaussian, curvature
perturbations predicted by inflation.
Until recently, $\Lambda$CDM's only major flaw was the absence
of evidence for accelerated expansion.  Not now.

\section{Three Dark Matter Problems}

While stars are very interesting and pretty to look at --
and without them, astronomy wouldn't be astronomy and we
won't exist -- they
represent a tiny fraction of the cosmic mass budget, only about 0.5\%
of the critical density.  As we have know for several decades
at least -- the bulk of the matter and energy in the Universe
is dark.  The present accounting defines clearly three dark matter/energy
problems; none is presently fully addressed.

\subsection{Dark Baryons}

By a ten to one margin, the bulk of the baryons do not exist in
the form of bright stars.  With the exception of clusters, where the
dark baryons exist in the form of hot, x-ray emitting
intracluster gas, the nature of the dark baryons is not known.
Clusters of course only account for around 10\% or so of the
matter in the Universe and the (optically) dark baryons elsewhere
could take on a different form.

The two most promising possibilities for the dark baryons are
diffuse hot gas and ``dark stars''
(white dwarfs, neutron stars, black holes or objects of mass around
or below the hydrogen-burning limit).  The former
possibility is favored by me for a number of reasons.  First, that's
where the dark baryons in clusters are.  Second, the cluster baryon
fraction argument can be turned around to infer $\Omega_{\rm gas}$
at the time clusters formed, redshifts $z\sim 0 -1$,
\begin{equation}
\Omega_{\rm gas}h^2 = f_{\rm gas} \Omega_Mh^2 =
0.023\,(\Omega_M/0.4)(h/0.65)^{1/2}\,.
\end{equation}
That is, at the time clusters formed, the mean gas density was essentially
equal to the baryon density (unless $\Omega_Mh^{1/2}$ is very small),
thereby accounting for the bulk of baryons in gaseous form.
Third, numerical simulations suggest that most of
the baryons should still be in gaseous form today (Rauch et al, 1997).

I should mention that there are two arguments for dark stars as the
baryonic dark matter.  First,
the gaseous baryons not associated with clusters have not been
detected.  Second, the results
of the microlensing surveys (see Alcock, 1999)
toward the LMC and SMC are consistent
with about one-third of our halo being in the form of half-solar
mass white dwarfs.

I find neither argument compelling;
gas outside clusters will be much cooler ($T\sim 10^5 - 10^6$\,K)
and very difficult to detect, either in absorption or emission.
There are equally attractive explanations for the Magellanic
Cloud microlensing events (e.g., self lensing by the Magellanic
Clouds, lensing by stars in the spheroid, or lensing due to
disk material that, due to flaring and warping of the disk,
falls along the line of sight to the LMC; see Sahu, 1994;
Evans et al, 1998; Gates et al, 1998; Zaritsky \& Lin, 1997;
Zhao, 1998).
The white-dwarf interpretation for the halo has a host of troubles:
Why haven't the white dwarfs been seen (Graff et al, 1998)?  The
star formation rate required to produce these white dwarfs --
close to $100\,{\rm yr^{-1}\,Mpc^{-3} }$ -- far exceeds
that measured for other parts of the Universe.  Where are the
lower-main-sequence stars associated with this stellar
population and the gas (expected to be 6 to 10 times that
of the white dwarfs) that didn't form into stars (Fields et al,
1997)?  Finally, there is evidence that the
lenses for both SMC events are stars within the SMC (Alcock et al,
1998; EROS Collaboration, 1998a,b)
and at least one of the LMC events is explained by an LMC lens.

\subsection{Cold Dark Matter}

The second dark-matter problem follows from the inequality
$\Omega_M\simeq 0.4 \gg \Omega_B\simeq 0.05$:
There is much more matter than
there are baryons, and thus, nonbaryonic dark matter is
the required, dominant form of matter.  The evidence for this very
profound conclusion has been mounting for almost two decades.
This year, the Burles -- Tytler deuterium measurement anchored the
baryon density and allowed the cleanest determination of
the matter density.

Particle physics provides an attractive solution to the
nonbaryonic dark matter problem:  relic elementary particles
left over from the big bang.  Long-lived or stable particles
with very weak interactions can remain from the earliest
moments of particle democracy in sufficient numbers to account
for a significant fraction of critical density (very weak interactions
are needed so that their annihilations cease before their
numbers are too small).  The three most promising candidates
are a neutrino(s) of mass 30\,eV or so, an axion of mass
$10^{-5\pm 1}\,$eV, and a neutralino of mass between $50\,$GeV and $500\,$GeV.
All three are motivated by particle physics theories that attempt
to unify the forces and particles of Nature.  The fact that
such particles can also account for the nonbaryonic dark matter
is either a big coincidence or a big hint.  Further, the fact
that these particles interact with each other and ordinary very
weakly, provides a simple and natural explanation for dark matter
being more diffusely distributed.

At the moment, there is significant circumstantial
evidence against neutrinos as the bulk of the dark matter.  Because
they behave as hot dark matter, structure forms from
the top down, with superclusters fragmenting into clusters
and galaxies (White, Frenk \& Davis, 1983),
in stark contrast to the observational
evidence that indicates structure formed from the bottom
up.  (Hot + cold dark matter is still an outside possibility,
with $\Omega_\nu \sim 0.15$; see Gawiser \& Silk, 1998.)
Second, the evidence for neutrino mass based upon
the atmospheric- and solar-neutrino data suggests a
neutrino mass pattern with the tau neutrino at $0.1\,$eV,
the muon neutrino at $0.001\,$eV to $0.01$\,eV and the
electron neutrino with an even smaller mass.  In particular,
the factor-of-two deficit of atmospheric muons neutrinos with
its dependence upon zenith angle is very strong evidence
for a neutrino mass difference
squared between two of the neutrinos of around $10^{-2}$\,eV$^2$
(Fukuda et al, 1998).  In turn, this sets a lower bound to
neutrino mass of about $0.1\,$eV, implying neutrinos contribute
at least as much mass as bright stars.  WOW!

Both the axion and neutralino behave as cold dark matter; the
success of the cold dark matter model of structure formation
makes them the leading particle dark-matter candidates.  Because
they behave as cold dark matter, they are expected to be the
dark matter in our own halo -- in fact, there is nothing that
can keep them out (Gates \& Turner, 1994).  As discussed above,
2/3 of the dark matter in our halo -- and probably all the
halo dark matter -- cannot be explained by baryons in any form.
The local density of halo material is estimated to be $10^{-24}\,{\rm g\,
cm^{-3}}$, with an uncertainty of slightly less than a factor
of 2 (Gates et al, 1995).  This makes the halo of our galaxy
an ideal place to look for cold dark matter particles!
An experiment at Livermore
National Laboratory with sufficient sensitivity to detect halo
axions is currently taking data (van Bibber et al,
1998) and experiments at several laboratories around the
world are beginning to search for halo neutralinos with sufficient sensitivity
to detect them (Sadoulet, 1999).
The particle dark matter hypothesis is a
very bold one, and it is now being tested.

\subsection{Dark Energy}

I and others have often used the term exotic to refer to particle
dark matter.  That term will now have to be reserved for the
dark energy that is causing the accelerated expansion of the
Universe -- by any standard, it is more exotic and more
poorly understood.

Here is what we do know:   it contributes
about 60\% of the critical density
and has pressure more negative than $-\rho /3$ (i.e., effective
equation of state $w\equiv p/\rho < - {1\over 3}$).  It does
not clump (otherwise it would have contributed to estimates
of the mass density).  The simplest possibility is that it
is the energy associated with the virtual particles that populate
the quantum vacuum (which has equation of state $w=-1$ and
is absolutely spatially uniform).

This simple interpretation has its difficulties.   Einstein ``invented''
the cosmological constant to make a static model of the Universe
and then he discarded it; we now know that the concept is not optional.
The cosmological constant corresponds to the energy associated
with the vacuum.  However, there is no sensible calculation of
that energy (see e.g., Weinberg, 1989),
with estimates ranging from $10^{122}$ to $10^{55}$ times the critical
density.  Some particle physicists believe that when the
problem is understood, the answer will be zero.  Spurred
in part by the possibility that cosmologists may have actually weighed
the vacuum (!), particle theorists are taking a fresh look
at the problem (see e.g., Harvey, 1998; Sundrum, 1997).  Sundrum's
proposal, that the energy of the vacuum is close to the present
critical density because the graviton is a composite particle
with size of order 1\,cm, is indicative of the profound
consequences that a cosmological constant has for fundamental physics.

Because of the theoretical problems mentioned above, as well as the
checkered history of the cosmological constant, theorists have
explored other possibilities for a smooth, component to the dark
energy (see e.g., Turner \& White, 1997).
Wilczek and I pointed out that even if the
energy of the true vacuum is zero, as the Universe as
cooled and went through a series of phase transitions, it
could have become hung up in a metastable vacuum with
nonzero vacuum energy (Turner \& Wilczek, 1982).  In the
context of string theory, where there are a very large number
of energy-equivalent vacua this becomes a more
interesting possibility:  perhaps the degeneracy of vacuum
states is broken by very small effects, so small that
we were not steered into the lowest energy vacuum during
the earliest moments.

Vilenkin (1984) has suggested a tangled network of very light
cosmic strings (also see, Spergel \& Pen, 1997) produced
at the electroweak phase transition; networks of other frustrated
defects (e.g., walls) are also possible.  In general, the
bulk equation-of-state of frustrated defects
is characterized by $w=-N/3$ where $N$
is the dimension of the defect ($N=1$ for strings, $=2$
for walls, etc.).  The SNeIa data almost exclude strings,
but still allow walls.

An alternative that has received a lot of attention is
the idea of a ``decaying cosmological constant'', a termed
coined by the Soviet cosmologist Matvei Petrovich Bronstein in 1933
(Bronstein, 1933).  (Bronstein was executed on Stalin's
orders in 1938, for reasons not directly related to the
cosmological constant.)
The term is, of course, an oxymoron; what people have in mind
is making vacuum energy dynamical.  The simplest realization
is an evolving scalar field.  If it is spatially homogeneous,
then its energy density and pressure are given by
\begin{eqnarray}
\rho & = & {1\over 2}{\dot\phi}^2 + V(\phi ) \nonumber \\
p    & = & {1\over 2}{\dot\phi}^2 - V(\phi )
\end{eqnarray}
and its equation of motion by (see e.g., Turner, 1983)
\begin{equation}
\ddot \phi + 3H\dot\phi + V^\prime (\phi ) = 0
\end{equation}

The basic idea is that energy of the true vacuum is zero, but
not all fields have evolved to their state of minimum
energy.  This is qualitatively different from that
of a metastable vacuum, which is a local minimum of the
potential and is classical stable.  Here, the field is
classically unstable and is rolling toward its lowest
energy state.

Two features of the ``rolling-scalar-field scenario'' are
worth noting.  First, the effective equation of state,
$w=({1\over 2}\dot\phi^2 - V)/({1\over 2}\dot\phi^2 +V)$,
can take on any value from 1 to $-1$.  Second, $w$ can
vary with time.  These are key features that allow it
to be distinguished from the other possibilities.

The rolling scalar field scenario (aka mini-inflation
or quintessence) has received a lot of attention over
the past decade (Freese et al, 1987; Ozer \& Taha, 1987;
Ratra \& Peebles, 1988; Frieman et al, 1995; Coble et al, 1996;
Turner \& White, 1997; Caldwell et al, 1998).  It is an interesting idea,
but not without its own difficulties.  First, in this
scenario one must {\em assume} that the energy of the
true vacuum state ($\phi$ at the minimum of its potential)
is zero; i.e., it does not address the cosmological
constant problem.  Second, as Carroll (1998) has emphasized,
the scalar field $\phi$ is very light and can
mediate long-range forces.
This places very stringent constraints on it.  Finally,
with the possible exception of one model, none of the
scalar-field models address how $\phi$ fits into the
grander scheme of things and why it is so light ($m\sim 10^{-33}\,$eV).

\section{Concluding Remarks}

1998 was a very good year for cosmology.  We have
for the first time a plausible, complete accounting
of matter and energy in the Universe, in $\Lambda$CDM
a model for structure formation that is consistent with
all the data at hand, and the first evidence for the key
tenets of a bold and expansive paradigm that extends the
standard hot big-bang model (Inflation + Cold Dark Matter).
One normally conservative cosmologist has gone out on
a limb by saying that 1998 may be a turning point in
cosmology as important as 1964, when the CBR was discovered
(Turner, 1999)!

We still have important questions to address:  Where
are the dark baryons?  What is the dark matter?  What
is the nature of the dark energy?  What is the explanation for the
complicated pattern of mass and energy:
neutrinos (0.3\%), baryons (5\%),
cold dark matter particles (35\%) and dark energy (60\%)?
Especially puzzling is the ratio of dark energy to dark matter:
because they evolve differently, the ratio of dark matter
to dark energy was higher in the past and will be smaller in
the future; only today are they comparable.  WHY NOW?

While we have many urgent questions, we can see
a flood of precision cosmological and laboratory
data coming that will help to answer these questions:
High-resolution maps of CBR anisotropy (MAP and Planck);
large redshift surveys (SDSS and 2dF); more SNeIa data;
experiments to directly detect halo axions and neutralinos;
more microlensing data (MACHO, EROSII, OGLE, AGAPE, and
superMACHO); accelerator experiments at Fermilab and CERN
searching for the neutralino and its supersymmetric friends
and further evidence for neutrino mass; and
nonaccelerator experiments that will shed further light
on neutrino masses, particle dark matter, new forces,
and the nature of gravity.

These are exciting times in cosmology!

\acknowledgments
This work was supported by
the DoE (at Chicago and Fermilab) and by the NASA (at Fermilab
by grant NAG 5-7092).


\end{document}